# Correlated connectivity and the distribution of firing rates in the neocortex

Alexei Koulakov, Tomas Hromadka, and Anthony M. Zador

*Cold Spring Harbor Laboratory, Cold Spring Harbor, NY 11724*

**ABSTRACT**: Two recent experimental observations pose a challenge to many cortical models. First, the activity in the auditory cortex is sparse, and firing rates can be described by a lognormal distribution. Second, the distribution of non-zero synaptic strengths between nearby cortical neurons can also be described by a lognormal distribution. Here we use a simple model of cortical activity to reconcile these observations. The model makes the experimentally testable prediction that synaptic efficacies onto a given cortical neuron are statistically correlated, *i.e.* it predicts that some neurons receive many more strong connections than other neurons. We propose a simple Hebb-like learning rule which gives rise to both lognormal firing rates and synaptic efficacies. Our results represent a first step toward reconciling sparse activity and sparse connectivity in cortical networks.

## Introduction

The input to any one cortical neuron consists largely of the output from other cortical cells (Benshalom and White, 1986; Douglas et al., 1995; Suarez et al., 1995; Stratford et al., 1996; Lubke et al., 2000). This simple observation, combined with experimental measurements of cortical activity, impose powerful constraint on models of a cortical circuits. The activity of any cortical neuron selected at random must be consistent with that of the other neurons in the circuit. Violations of self-consistency pose a challenge with theoretical models of cortical networks.

A classic example of such a violation was the observation (Softky and Koch, 1993) that the irregular Poisson-like firing of cortical neurons is inconsistent with a model in which each neuron received a large number of uncorrelated inputs from other cortical neurons firing irregularly. Many resolutions of this apparent paradox were subsequently proposed (van Vreeswijk and Sompolinsky, 1996; Troyer and Miller, 1997; Shadlen and Newsome, 1998; Salinas and Sejnowski, 2002). One resolution (Stevens and Zador, 1998)—that cortical firing is not uncorrelated, but is instead organized into synchronous volleys, or "bumps"—was recently confirmed experimentally in the auditory cortex (DeWeese and Zador, 2006). Thus a successful model can motivate new experiments.

Two recent experimental observations pose a new challenge to many cortical models. First, it has recently been shown (Hromadka et al., 2008) that activity in the primary auditory cortex of awake rodents is sparse. Specifically, the distribution of spontaneous firing rates can be described by a lognormal distribution (Figure 1A and B). Second, the distribution of non-zero synaptic strengths measured between pairs of connected cortical neurons is also well-described by a lognormal distribution (Figure 1C and D; (Song et al., 2005)). As shown below, the simplest randomly connected model circuit that incorporates a lognormal distribution of synaptic weights predicts that firing rates measured across the population will have a Gaussian rather than a lognormal distribution. The observed lognormal distribution of firing rates therefore imposes additional constraints on cortical circuits.

In this paper we address two questions. First, how can the observed lognormal distribution of firing rates be reconciled with the lognormal distribution of synaptic efficacies? We find that reconciling lognormal firing rates and synaptic efficacies implies that inputs onto a given cortical neuron must be statistically correlated—an experimentally testable prediction. Second, how might the distributions of emerge in development? We propose a simple Hebb-like learning rule which gives rise to both lognormal firing rates and synaptic efficacies.

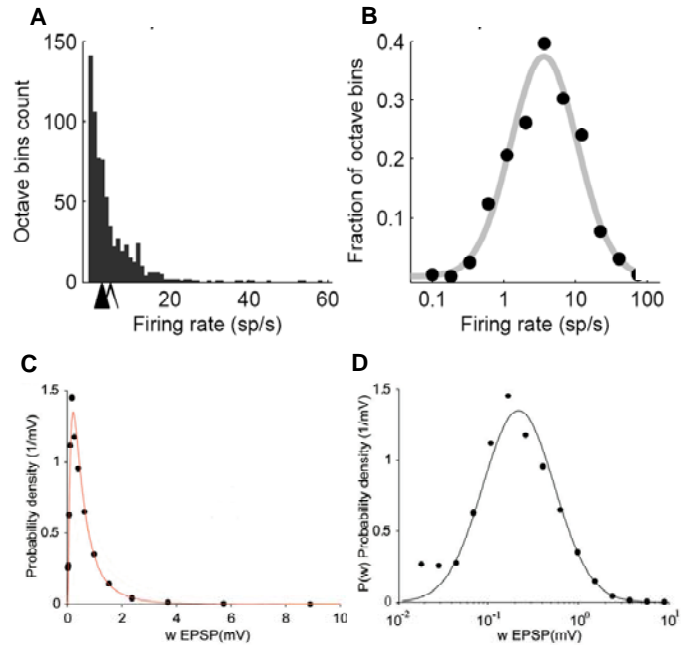

**Figure 1.** Lognormal distributions in cerebral cortex. (A, B) Distribution of spontaneous firing rates in auditory cortex of unanesthetized rats follows a lognormal distribution (Hromadka et al., 2008). Measurements with the cell-attached method show that spontaneous firing rates in cortex vary within several orders of magnitude.

The distribution is fit well by a lognormal distribution with some cells displaying values of firing rate above 30 Hz and an average firing rate of about 3 Hz (black arrow).

(C, D) The distribution of synaptic weights for intracortical connections (Song et al., 2005). To assess this distribution, pairs of neurons in the network were chosen randomly and the strength of the connections between them is measured using electrophysiological methods (Song et al., 2005). Most of connections between pairs turns out to be of zero strength: the sparseness of cortical network is about 20% even if the neuronal cell bodies are close to each other (Stepanyants et al., 2002). This implies that in about 80% of pairs there is no direct synaptic connection. The distribution of non-zero synaptic efficacies is close to lognormal (Song et al., 2005), at least, for the connectivity between neurons in layer V of rat visual cortex. This implies that the logarithm of the synaptic strength has a normal (Gaussian) distribution.

## Methods

### Generation of lognormal matrices

Here we describe the methods used for generating weight matrices in Figures 2—4. These matrices were constructed using the MATLAB random number generator. Figure 2 displays a purely white-noise matrix with no correlations between elements. To generate the lognormal distribution of the elements of this matrix we first generated a matrix $\hat{N}$ whose elements are distributed normally, with zero mean and a unit standard deviation. The white-noise weight matrix $\hat{W}$ was then obtained by evaluating exponential of the individual elements of $\hat{N}$, i.e. $W_{ij} = \exp(N_{ij})$. Elements of the weight matrix obtained with this method have a lognormal distribution since their logarithms ($N_{ij}$) are normal. To obtain the column-matrix (Figure 3) we used the following property of the lognormal distribution: The product of two lognormally distributed numbers is also lognormally distributed. The column matrix can therefore be obtained by multiplying the columns of a white-noise lognormal matrix $A_{ij}$, which is generated using the method described above, by a set of lognormal numbers $v_j$, i.e.

$$W_{ij} = A_{ij} v_j. \qquad (1)$$

Both $A_{ij}$ and $v_j$ have zero mean and a unit standard deviation. Similarly, the row-matrix in Figure 4 is obtained by multiplying each row of the white-noise matrix $A_{ij}$ with the set of numbers $v_i$:

$$W_{ij} = v_i A_{ij} \qquad (2)$$

As in equation (1) both $A_{ij}$ and $v_j$ are lognormally distributed with zero mean and unit standard deviation.

## Lognormal firing rates for row-matrices

Here we explain why the elements of the principal eigenvector of row-matrices have a broad lognormal distribution (Figure 4D). Consider the eigenvalue problem for the row-matrix represented by equation (10). It is described by

$$\sum_{j=1}^{N} v_i A_{ij} f_j = f_i \qquad (3)$$

Equation (3) can be rewritten in the following way

$$\sum_{j=1}^{N} A_{ij} v_j \frac{f_j}{v_j} = \frac{f_i}{v_i} \qquad (4)$$

Thus the vector $y_i = f_i / v_i$ is the eigenvector of the column-matrix $A_{ij} v_j$ [cf. equation (1)]. As such, it is a normally distributed quantity with low CV as shown in Figure 3.

$$y_i \approx 1 \qquad (5)$$

This approximate equality becomes more precise as the size of the weight matrix goes to infinity. Therefore we conclude that

$$f_i \approx v_i. \qquad (6)$$

Because $A_{ij}$ and $v_j$ are lognormal, both $W_{ij} = v_i A_{ij}$ and its eigenvector $f_i \approx v_i$ are also lognormal.

## Non-linear learning rule

We will demonstrate here that the non-linear Hebbian learning rule given by equation (11) can yield row-matrix as described by equation (2) in the state of equilibrium. Because of the requirement of equilibrium we can assign $\dot{W}_{ij} = 0$ after what equation (11) yields

$$W_{ij} = \left( \frac{\varepsilon_1}{\varepsilon_2} \right)^{\frac{1}{1-\beta}} f_i^{\frac{\alpha}{1-\beta}} f_j^{\frac{\alpha}{1-\beta}} C_{ij} \qquad (7)$$

Here $C_{ij}$ is the adjacency matrix (Figure 5B) whose elements are equal to either 0 or 1 depending on whether there is a synapse from neuron number $j$ to neuron $i$. Note that in this notation the adjacency matrix is transposed compared to the convention used in the graph theory. The firing rates of the neurons $f_i$ in the stationary equilibrium state are themselves components of the principal eigenvector of $W_{ij}$ as required by equation (10). After substituting equation (7) into equation (10) simple algebraic transformations lead to

$$f_i \sim \left( \sum_j C_{ij} f_j^{1-\beta} \right)^{\frac{\gamma}{1-\alpha-\beta}} \sim (1+\xi_i)^{\frac{1-\beta}{1-\alpha-\beta}} \qquad (8)$$

Because the elements of the adjacency matrix are uncorrelated in our model the sum in equation (8) has Gaussian distribution with small coefficient of variation vanishing in the limit of large network. Therefore the variable $\xi_i$ describing relative deviation of this sum for neuron $i$ from the mean is normal with variance much smaller than one. Taking the logarithm of equation (8) and taking advantage of the smallness of variance of $\xi_i$ we obtain

$$\ln f_i \approx \frac{1-\beta}{1-\alpha-\beta} \xi_i . \qquad (9)$$

Because $\xi_i$ is normal, $f_i$ is lognormal. This is confirmed by Figure 5B. In the limit $\alpha + \beta \to 1$ the variance of the lognormal distribution of $f_i$ diverges according to equation (9). Thus even if $\xi_i$ has small variance, firing rates may be broadly distributed with the standard deviation of its logarithm reaching unity as in Figures 5 and 7. The non-zero elements of the weight matrix are also lognormally distributed, because, according to equation (7) weight matrix is a product of powers of lognormal numbers $f_i$. These conclusions are discussed in more detail in the Supplementary Materials.

### Details of computer simulations

To generate Figures 5-7 we modeled the dynamics described by equation (11). The temporal derivatives were approximated by discrete differences $\dot{W}_{ij} \approx \Delta W_{ij} / \Delta t$ with the time step $\Delta t = 1$, as described in more detail in Supplementary Materials. The simulation included 1000 iterative steps. We verified that both of the distributions of the firing rates and weights saturates and stays approximately constant at the end of the simulation run. For every time step the distribution of spontaneous firing rates was calculated from equation (10) taking the elements of the principal eigenvector of matrix $W_{ij}$. Since the eigenvector is defined up to a constant factor, the vector of firing rates obtained this way was normalized to yield zero average logarithm of its elements. The weight matrix was also normalized by dividing it with the maximal eigenvalue, thus yielding the eigenvalue of one. These normalizations were performed on each step and were intended to mimic the homeostatic controls of the average firing rates and overall scale of synaptic weights in the network. A multiplicative noise of 5% was added to the vector of firing rates on each iteration step. The parameters used were $\alpha = \beta = 0.4$, $\gamma = 0.45$ in Figures 5 and 6, and $\alpha = \beta = 0.36$, $\gamma = 0.53$ in Figure 7. Before iterations started random adjacency matrices were generated with 20% sparseness (Figures 5B and 7B). These matrices contained 80% of zeros and 20% of elements that were either +1 or -1 depending on whether the connection is excitatory or inhibitory. In Figure 5 only excitatory connections were present. In Figure 7 the adjacency matrix contained 15% of 'inhibitory' columns representing axons of inhibitory neurons. In these columns all of the non-zero matrix elements were equal to -1. The weight matrices were initialized to the absolute value of the adjacency matrices before the evolution in time was simulated as described above.

### Results

### Recurrent model of spontaneous cortical activity

To model the spontaneous activity of the $i^{th}$ neuron in the cortex, we assume that its firing rate $f_i$ is given by a weighted sum of the firing rates $f_j$ of all the other neurons in the network:

$$f_i = \sum_{j=1}^{N} W_{ij} f_j. \qquad (10)$$

Here $W_{ij}$ is the strength of the synapse connecting neuron $j$ to neuron $i$. This expression is valid if the external inputs, such as thalamocortical projections, are weak (for example, in the absence of sensory inputs, when the spontaneous activity is usually measured), or when recurrent connections are strong enough to provide significant amplification of the thalamocortical inputs (Douglas et al., 1995; Suarez et al., 1995; Stratford et al., 1996; Lubke et al., 2000). Throughout this study we will use a linear model for the network dynamics, both because it is the simplest possible approach that captures the essence of the problem and because the introduction of non-linearity does not change our main conclusions (*see* Supplementary materials).

Equation (10) defines the consistency constraint between the spontaneous firing rates $f_j$ and the connection strengths $W_{ij}$ mentioned in the introduction. Indeed, given the weight matrix, not all values of spontaneous firing rates can satisfy this equation. Conversely, not any distribution of individual synaptic strengths (elements of matrix $W_{ij}$) is consistent with the particular distribution of spontaneous activities (elements of $f_j$). It can be recognized that equation (10) defines an eigenvector problem, a standard problem in linear algebra (Strang, 2003). Specifically, the set of spontaneous firing rates represented by vector $\vec{f}$ is the principal eigenvector (*i.e.* the eigenvector with the largest associated eigenvalue) of the connectivity matrix $\hat{W}$ (Rajan and Abbott, 2006). The eigenvalues and eigenvectors of a matrix can be determined numerically using a computer package such as MATLAB.

Before proceeding, we note an additional property of our model. In order for the principal eigenvector to be stable, the principal eigenvalue must be unity. If the principal eigenvector is greater than unity then the firing rates grow without bound to infinity, whereas if the principal eigenvalue is less than one the firing rates decay to zero. Mathematically, it is straightforward to renormalize the principal eigenvalue by considering a new matrix formed by dividing all the elements of the original matrix by its principal eigenvalue. Biologically such a normalization may be accomplished by global mechanisms controlling the overall scale of synaptic strengths, such as the homeostatic control (Davis, 2006), short-term synaptic plasticity, or synaptic scaling (Abbott and Nelson, 2000). Our model is applicable if any of the above mechanisms are at play.

*Recognizing that Equation* (10) *defines an eigenvector problem allows us to recast the first neurobiological problem posed in the introduction as a mathematical problem.* We began by asking whether it was possible to reconcile the observed lognormal distribution of firing rates (Figure 1A) with the observed lognormal distribution of synaptic efficacies (Figure 1B). Mathematically, the experimentally observed distribution of spontaneous firing rates corresponds to the distribution of the elements $f_i$ of the vector of spontaneous firing rates $\vec{f}$, and the experimentally observed distribution of synaptic efficacies corresponds to the distribution of non-zero elements $W_{ij}$ of the synaptic connectivity matrix $\hat{W}$. Thus the mathematical problem is: *Under what conditions does a matrix $\hat{W}$ whose non-zero elements $W_{ij}$ obey a lognormal distribution has a principal eigenvector $\vec{f}$ whose elements $f_i$ also obey a lognormal distribution?*

In the next sections we first consider synaptic matrices whose elements are non-negative numbers. Such synaptic matrices describe networks containing excitatory neurons in which zero connection strength implies simply that there is no synapse, while positive synaptic values describe synaptic efficacy between excitatory cells. The properties of the principal eigenvalues and eigenvectors of such matrices are described by the Perron-Frobenius theorem (Varga, 2000). This theorem ensures that the principal eigenvalue of the synaptic matrix is a positive real number, that there is only one solution for the principal eigenvalue and eigenvector, and that the elements of the eigenvector representing in our case spontaneous firing rates of individual neurons are all positive. These properties are valid for the so-called irreducible matrices which describe networks in which activity can travel between any two nodes (Varga, 2000). Because we will consider either fully connected or sparse networks with connectivity above the percolation threshold (Stauffer and Aharony, 1992; Henrichsen, 2000), our matrices are irreducible. Later we will include inhibitory neurons by making some of the matrix elements negative. Although the conclusions of the Perron-Frobenious theorem do not apply directly to these networks, we have found experimentally that they are still valid, perhaps because the fraction of inhibitory neurons was kept small in our model (see below).

## Randomly connected lognormal networks do not yield lognormal firing

We first examined the spontaneous rates produced by a synaptic matrix in which there are no correlations between elements. We call this form of connectivity "white-noise" (Figure 2A). Note that the values of synaptic strength in this matrix have a lognormal distribution (Figure 2B), as observed in the experiments measuring the distribution of pair-wise synaptic strengths in cortex (Figure 1A) (Song et al., 2005). The standard deviation of the natural logarithm of non-zero connectivity strengths was taken to be equal to one, consistent with the experimental observations. The distribution of the spontaneous firing rates, obtained by solving the eigenvector problem, are displayed in Figure 2D. The spontaneous firing rates had similar values for all cells in the network, with a coefficient of variation of about 5%. It is clear that this distribution is quite different from the experimentally observed (Figure 1), in which the rates varied over at least one order of magnitude.

To understand why the differences in the spontaneous firing rates between cells are not large with white noise connectivity, consider two cells in a network illustrated in Figure 2C by red and blue circles. Width of connecting edges is proportional to connection strength, and the circle diameters are proportional to firing rates. All inputs into the two marked cells come from the same distribution with the same mean. This is a property of the white-noise matrix. Since each cell receives a large number of such inputs, the differences in inputs between these two cells are small, due to the central limit thorem. The inputs are approximately equal to the mean values multiplied by the number of inputs. Therefore one should expect that the firing rates of the cells are similar, as observed in our computer simulations.

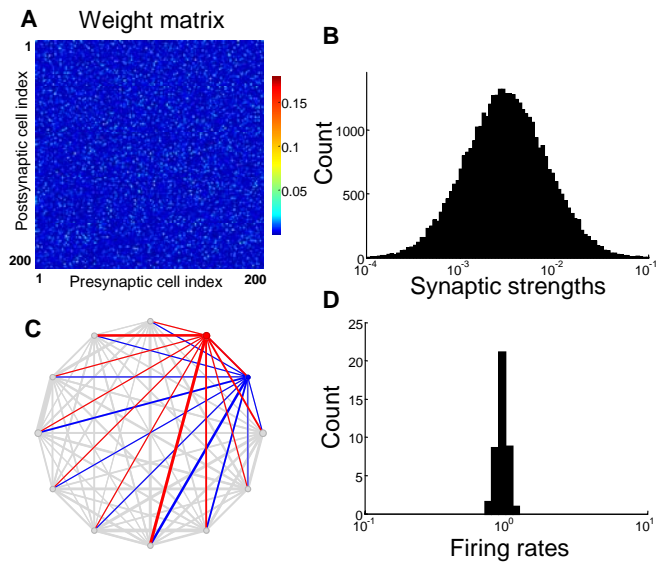

**Figure 2.** Randomly connected "white noise" network connectivity does not yield lognormal distribution of spontaneous firing rates.
(A) Synaptic connectivity matrix for 200 neurons. Because synaptic strengths are uncorrelated, the weight matrix looks like a "white-noise" matrix.
(B) Distribution of synaptic strengths is lognormal. The matrix is rescaled to yield a unit principal eigenvalue.

(C) Synaptic weights and firing rates of 12 randomly chosen neurons tended to be similar. Each circle corresponds to one neuron, with diameter proportional to its spontaneous firing rate. Thickness of connecting lines is proportional to strengths (synaptic weights) of incoming connections for each neuron. Red and blue circles and lines show spontaneous firing rates and incoming connection strengths for two neurons with maximum and minimum firing rates from the sample shown. Because incoming synaptic weights are similar on average the spontaneous firing rates (circle diameters) tend to be similar.

(D) Spontaneous firing rates given by the components of principal eigenvector of matrix shown in (A). The distribution of spontaneous firing rates in *not* lognormal, contrary to experimental findings (see Figure 1A and B). The spontaneous firing rates are approximately the same for all neurons in the network.

The connectivity matrix with no correlations between synaptic strengths therefore is inconsistent with experimental observations of dual lognormal distributions for both connectivity and spontaneous activity. We next explored the possibility that introducing correlations between connections would yield the two lognormal distributions.

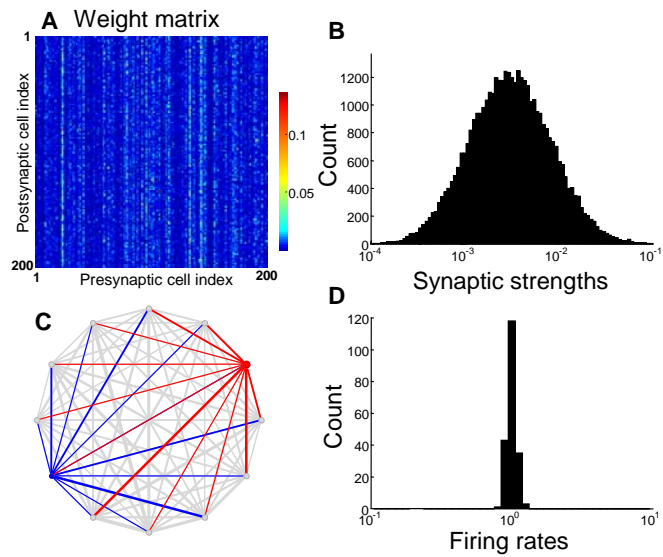

**Figure 3.** Correlated synaptic weights on the same axon (output correlations) do not lead to lognormal distribution of spontaneous firing rates.
(A) Synaptic weight matrix for 200 neurons contains vertical "stripes" indicating correlations between synapses made by the same presynaptic cell (the same axon).
(B) Distribution of synaptic weights is lognormal.
(C) Firing rates and synaptic weights tended to be similar for different neurons in the network, as illustrated on an example of 12 randomly chosen neurons. Red and blue circles show neurons with maximum and minimum firing rates (out of the sample shown), with their corresponding incoming connections.
(D) Column-matrix fails to yield broader distribution of spontaneous firing rates than the "white noise" matrix (Figure 2).

## Presynaptic correlations do not yield lognormal firing

We first considered the effect of correlations between the strengths of synapses made by a particular neuron. These synapses are arranged in the same column in the layout of the connectivity matrix shown in Figure 3A. This matrix is therefore denoted as a column-matrix. To create these correlations we generated a white-noise lognormal matrix and then multiplied each column by a random number chosen from another lognormal distribution. The elements of resulting column-matrix are also lognormally distributed (Figure 3B) as products of two lognormally distributed random numbers (see Methods).

As is clear from Figure 3, presynaptic correlations do not resolve the experimental paradox between the distributions of spontaneous firing rates and synaptic strengths. Although the connectivity matrix is lognormal (Figure 3B), the spontaneous activity has a distribution with low variance (Figure 3D). A different type of correlations is needed to explain high variances in both distributions.

The reason why the column-matrix fails to produce dual lognormal distributions is essentially the same as in the case of white-noise matrix. Each neuron in the network receives connections that are taken from the distributions with the same mean. When the number of inputs is large, the differences between inputs into individual cells become small due to the central limit theorem, with the total input being approximately equal to the average of the distribution multiplied by the number of inputs. Thus two cells in Figure 3C receive a large number of inputs with the same mean. There are correlations between inputs from the same cell (arrows) but these correlations only increase the similarity in firing between two cells. For this reason the variance of the distribution of the spontaneous firing rates is smaller in the case of column-matrix (Figure 3D) than in the case of white-noise connectivity (Figure 2D) as shown in the Supplementary Materials (Section 5). A different type of correlation is therefore needed to resolve the apparent paradox defined by the experimental observations.

**Postsynaptic correlations yield lognormal firing**

We finally tried the connectivity in which synapses onto the same postsynaptic neuron were positively correlated. Because such synapses impinge upon the same postsynaptic cell, they reside in the rows of the connectivity matrix (Figure 4A). The matrix was obtained by multiplying the elements of the white-noise matrix sharing the same row by the same number taken from the lognormal distribution (see Methods). This approach was similar to the generation of the column-matrix. It ensured that the non-zero synaptic strengths have a lognormal distribution (Figure 4B).

The resulting distribution of the spontaneous firing rates was broad (Figure 4D). It had all the properties of the lognormal distribution, such as the symmetric Gaussian histogram of the logarithms of the firing rates (Figure 4D) One can also prove that the distribution of spontaneous rates as defined by our model is lognormal for the substantially large row-correlated connectivity matrix (see Methods). We conclude that the row-matrix does have a property to generate the lognormal distribution of spontaneous firing rates.

The reason why the row-matrix yields a broad distribution of firing rates is illustrated in Figure 4C. Two different neurons (blue and red) each receive a large number of connections in this case. But these connections are multiplied by two different factors, each depending on the postsynaptic cell. This fact is shown in Figure 4C by differing thickness of lines entering two different cells. This implies that the average values of the strengths of the synapses onto this neuron are systematically different. Since both non-zero matrix elements and the spontaneous rates in this case have a lognormal distribution, the positive correlations between strengths of synapses on the same dendrite could underlie the dual lognormal distributions observed experimentally.

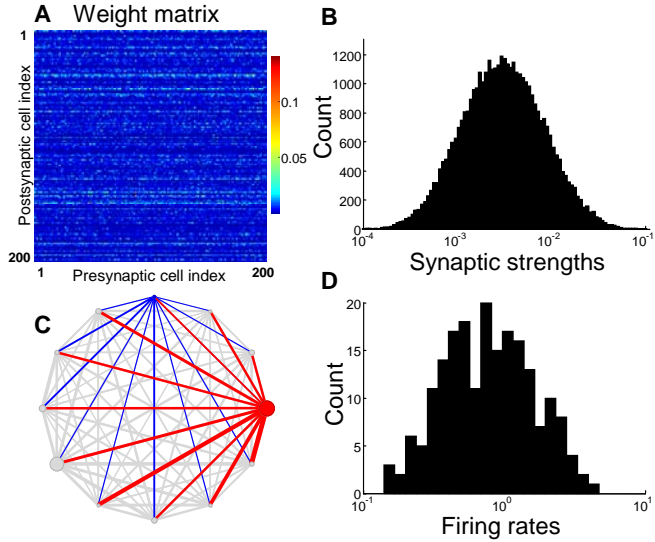

**Figure 4.** Correlations among synaptic weights on the same dendrite (input correlations) lead to lognormal distribution of spontaneous firing rates.
(A) Synaptic connectivity matrix for 200 neurons. Note the horizontal "stripes" showing input correlations.
(B) Distribution of synaptic weights is set up to be lognormal.
(C) Inputs into two cells, red and blue are shown by the thickness of lines in this representation of the network. Because synaptic strengths are correlated for the same postsynaptic cell, the inputs into cells marked by blue and red are systematically different, leading to large differences in the firing rates. For the randomly chosen subset containing 12 neurons shown in this example the spontaneous firing rates (circle diameter) vary widely due to large variance in the strength of incoming connections (line widths).
(D) Distribution of spontaneous firing rates is lognormal and has a large variance for row-matrix.

## Hebbian learning rule may yield lognormal firing rates and synaptic weights

In the previous section we showed that certain correlations in the synaptic matrix could yield lognormal distribution for spontaneous firing rates given lognormal synaptic strengths. A sufficient condition for this to occur is that the strengths of the synapses onto a given postsynaptic neuron must be correlated. To prove this statement we used networks that were produced by a random number generator (see Methods). The spontaneous activity then was the product of predetermined network connectivity. The natural question is whether the required correlations in connectivity can emerge naturally in the network through one of the known mechanisms of learning, such as Hebbian plasticity. Since Hebbian mechanisms strengthen synapses that have correlated activity, the synaptic connections become products of spontaneous rates too. Thus, network activity and connectivity are involved into mutually-dependent iterative process of modification. It is therefore not immediately clear if the required correlations in the network circuitry (row-matrix) can emerge from such an iterative process.

Rules for changing synaptic strength (learning rules) define the dynamics by which synaptic strengths change as a function of neural activity. We use the symbol $\dot{W}_{ij}$ to describe the rate of change in synaptic strength from cell number $j$ to $i$. In the spirit of Hebbian mechanisms, we assume that this rate depends on the presynaptic and postsynaptic firing rates, denoted by $f_j$ and $f_i$ respectively. In our model, in contrast to conventional Hebbian mechanism, it is also determined by the value of synaptic strength $W_{ij}$ itself, i.e.

$$\dot{W}_{ij} = \varepsilon_1 f_i^\alpha W_{ij}^\beta f_j^\gamma - \varepsilon_2 W_{ij} \qquad (11)$$

where as above $f_i$ and $f_j$ are firing rates of the post- and presynaptic neurons $i$ and $j$, respectively, and $\varepsilon_1$, $\varepsilon_2$, $\alpha$, $\beta$, and $\gamma$ are parameters discussed below. This equation implies that the rate of synaptic modification is a result of two processes: one for synaptic growth (the first term on the right hand side) and another for synaptic decay (the second term). The former process implements Hebbian potentiation, while the latter represents a passive decay. The relative strengths of these processes are determined by the parameters $\varepsilon_1$ and $\varepsilon_2$.

The Hebbian component is proportional to the product of pre- and postsynaptic firing rates and the current value of synaptic strength. Each of these factors is taken with some powers $\alpha$, $\beta$, $\gamma$, which are essential parameters of our model. When the sum of exponents $\alpha + \beta$ exceeds 1 a single weight dominates the weight matrix. The sum $\alpha+\beta$ of the exponents must be below 1 to prevent the emergence of winner-takes-it-all solutions. The learning rule considered here is therefore essentially non-linear.

When the sum of exponents $\alpha + \beta$ approaches 1 from below, the distribution of synaptic weights becomes close to lognormal. In the Methods section we prove this result. Here we present the results of computer simulation that illustrates this statement (Figure 5). The sum of exponents in this simulation is $\alpha + \beta = 0.8$, i.e. is very close to unity.

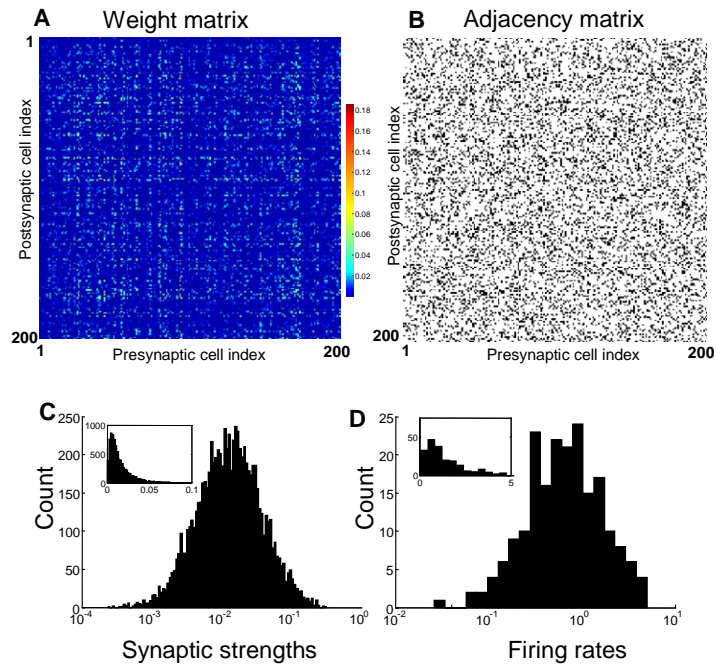

**Figure 5.**
Multiplicative Hebbian learning rule leads to lognormal network connectivity and firing rate distributions.
(A) Synaptic connectivity matrix (200 neurons) with "plaid" structure (horizontal and vertical "stripes"), similar to both column- and row-matrices introduced in previous sections. This matrix arose after 1000 iterations of multiplicative Hebbian learning rule (see text for details).
(B) The adjacency matrix[1] for the weight matrix is shown by an image in which existing/missing connections are black/white. Both weight and adjacency matrices are 20% sparse. The adjacency matrix is not symmetric, i.e. synaptic connections formed a directed graph. (C), (D) Distributions of synaptic weights (C) and firing rates (D) were lognormal, i.e. appeared as normally distributed on logarithmic axis.

---

[1] Note that adjacency defined here is transposed compared to the standard definition in graph theory.

In addition to a lognormal distribution of weights, the learning rule yields a lognormal distribution of spontaneous firing rates (Figure 5D). When the structure of synaptic matrix is examined visually, it reveals both vertical and horizontal correlations (Figure 5A). The resulting weight matrix therefore combines the features of row- and column- matrices. The lognormal distribution of spontaneous rates arises, as discussed above (Figure 4), from the correlations between inputs into each cell, *i.e.* from the row-structure of the synaptic connectivity matrix. The correlations between outputs (column-structure) emerge as a byproduct of the learning rule considered here. Because of the combined row-column correlations we call this type of connectivity patterns a "plaid" connectivity.

Although the matrix appear to be symmetric with respect to its diagonal (Figure 5A) the connectivity is not fully symmetric as shown by the distribution of non-zero elements in Figure 5B. It is notable that the learning rules used in this section [equation (11)] preserve the adjacency matrix. This implies that if two cells were not connected by a synapse, they will not become connected as a result of the learning rules. Similarly, synapses are not eliminated by the learning rule. Our Hebbian plasticity therefore preserves the sparseness of connectivity. In the Methods section we analyze the properties of plaid connectivity in greater detail. We conclude that multiplicative non-linear learning rule can produce correlations sufficient to yield dual lognormal distributions.

## Experimental predictions

Here we outline mathematical methods for detecting experimentally the correlations predicted by our model. Our basic findings are summarized in Figure 6A. For the lognormal distributions of both synaptic strength and firing rates (dual lognormal distributions) it is sufficient that the synapses of the same dendrite are correlated. This implies that the average strengths estimated for individual dendrites are broadly distributed. Thus, the synapses of the right dendrite in Figure 6A are stronger on average than the synapses on the left dendrite. This feature is indicative of the row-matrix correlations shown in Figures 6 and 5. In addition, if the Hebbian learning mechanism proposed here is implemented, the axons of the same cells should display a similar property. This implies that the average synaptic strength of each axon is broadly distributed. We suggest that these signatures of our theory could be detected experimentally.

Modern imaging techniques permit measuring synaptic strengths of substantial number of synapses localized on individual cells (Kopec et al., 2006; Micheva and Smith, 2007). These methods allow monitoring the postsynaptic indicators of connection strength in a substantial fraction of synapses belonging to individual cells. Therefore these methods could allow detecting the row-matrix connectivity (Figure 4) using the statistical procedure described below. The same statistical procedure could be applied to presynaptic measures of synaptic strengths to reveal plaid connectivity (Figure 5).

We will illustrate our method on the example of postsynaptic indicators. Assume that the synaptic strengths are available for several dendrites in the volume. First, for each cell we calculate the logarithm of average synaptic strength (LASS). We obtain a set of LASS characteristics matching in size the number of cells available. Second, the distribution of LASS is studied. The distribution for the row-matrix connectivity is wide, wider than expected for the white-noise matrix (Figure 6B, red histogram). A useful measure of the width of distribution is its standard deviation. For the dataset produced by the Hebbian learning rule used in the previous section the width of distribution of LASS is about 0.64 natural logarithm units (gray arrow in Figure 6C). Third, we assess the probability that the same width of distribution can be produced by the white-noise matrix, i.e. with no correlations present. To this end we employ a bootstrap procedure (Hogg et al., 2005). In the spirit of bootstrap we generate the white noise matrix from the data by randomly moving the synapses from dendrite to dendrite, either with or without repetitions. The random repositioning of the synapses preserves the distribution of synaptic strength but destroys the sought correlations, if they are present. The distribution of LASS is evaluated for each random repositioning of synapses of dendrites (iteration of bootstrap). One such distribution is shown for the data in the previous section in Figure 6B (black). It is clearly narrower than in the original dataset. By repeating the repositioning of synapses several types one can calculate the fraction of cases in which the width of the LASS distribution in the original dataset is smaller than the width in the reshuffled dataset. Smallness of this fraction implies that the postsynaptic connectivity is substantially different from the white-noise matrix. For the connectivity obtained by the Hebbian mechanism in the previous section, after $10^6$ iterations of bootstrap we observed *none* with the width of distribution of LASS larger than in the original non-permuted dataset (Figure 6C). We conclude that it is highly unlikely that the data in Figure 5 describe the white-noise matrix (p-value $< 10^{-6}$).

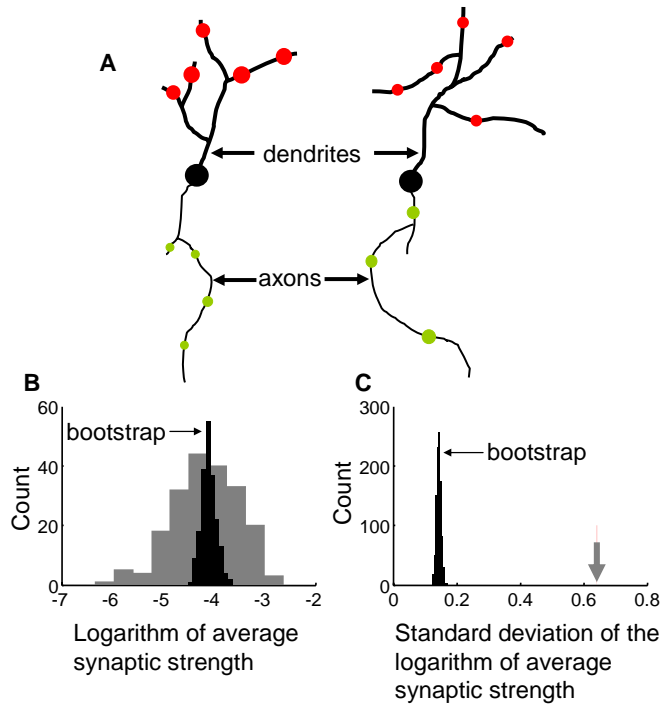

**Figure 6.**
Experimental predictions of this theory.
**(A)** The presence of row connectivity (Figure 4—5), sufficient for generation of dual lognormal distributions, implies correlations between synaptic strengths on each dendrite (the diameter of the red circle). In addition, if the non-linear Hebbian mechanism is involved in generation of these correlations, the synapses on the same axon are expected to be correlated (plaid-connectivity, Figure 5).
**(B)** To reveal these correlations, the logarithm of average synaptic strengths (LASS) was calculated for each dendrite. The distribution of these averages for dendrites (rows) from Figure 5 is shown by red bars. The standard deviation of this distribution is about 0.64 in natural logarithm units. The black histogram shows LASS distribution after the synapses were "scrambled" randomly, with their identification with particular dendrites removed. This bootstrapping procedure (Hogg et al., 2005) builds a white-noise matrix with the same distribution of synaptic weights, but much narrower distribution of bootstrapped LASS.
**(C)** Distribution of standard deviations (distribution widths) of LASS for many iterations of bootstrap (black bars). The widths were significantly lower than the width of the original LASS distribution (0.64, gray arrow). This feature is indicative of input correlations.

A similar bootstrap analysis could be applied to axons, if sets of synaptic strengths are measured for several axons in the same volume. A small p-value in this case would indicate the presence of column-matrix. The latter may be a consequence of the non-linear Hebbian mechanism proposed in the previous section.

## Inhibitory neurons

Cortical networks consist of a mixture of excitatory and inhibitory neurons. We therefore tested the effects of inhibitory neurons on our conclusions. We added a small (15%) fraction of inhibitory elements to our network. Introduction of inhibitory elements was accomplished through the use of an adjacency matrix. The adjacency matrix in this case described both the presence of a connection between neurons and the connection sign. Thus an excitatory synapse from neuron $j$ to neuron $i$ is denoted by an entry in the adjacency matrix $C_{ij}$ equal to one; inhibitory/missing synapses are described by entries equal to -1 or 0 respectively (Figure 7B). The presence of inhibitory neurons is reflected by the vertical column structure in the adjacency matrix (Figure 7B). Each blue column in Figure 7B represents the axon of a single inhibitory neuron. We then assumed that the learning rules described by equation (11) apply to the *absolute values* of synaptic strengths of both inhibitory and excitatory synapses with $W_{ij}$ defining the absolute value of synaptic strength, and the adjacency matrix $C_{ij}$ its sign. The synaptic strengths and spontaneous firing rate distributions are presented in Figure 7C, D after a stationary state was reached as a result of the learning rule (11). Both distributions are close to lognormal. In addition the synaptic matrix $W_{ij}$ displayed the characteristic plaid structure obtained by is previously for purely excitatory networks (Figure 5). We conclude that the presence of inhibitory neurons does not change our previous conclusions qualitatively.

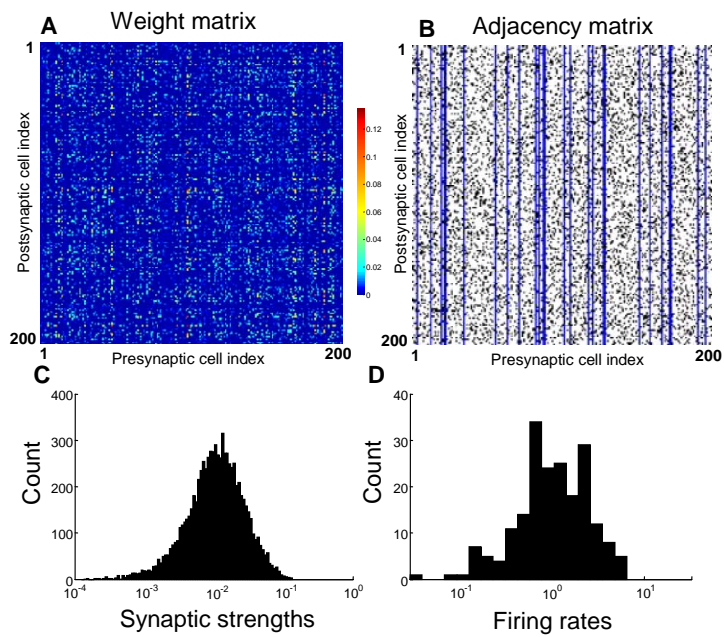

**Figure 7.** The results of non-linear multiplicative learning rule when inhibitory neurons are present in the network. (**A**) The absolute values of the weight matrix display the same 'plaid' correlations as in Figure 5A. (**B**) The adjacency matrix contains inhibitory connections. The presence of connection is shown by black points (20% sparseness). Inhibitory neurons are indicated by the vertical blue lines (15%). (**C**) The distribution of absolute values of synaptic strengths (also shown in A) is close to lognormal with small asymmetry. (**D**) The spontaneous firing rates are distributed approximately lognormally for this network containing inhibitory connections.

# Discussion

We have presented a simple model of cortical activity to reconcile the experimental observation that both spontaneous firing rates and synaptic efficacies in the cortex can be described by a lognormal distribution. We formulate this problem mathematically in terms of the distribution eigenvalues of the network connectivity matrix. We show that the two observations can be reconciled if the connectivity matrix has a special structure; this structure implies that some neurons receive many more strong connections than other neurons. Finally, we propose a simple Hebb-like learning rule which gives rise to both lognormal firing rates and synaptic efficacies.

## Lognormal distributions in the brain

The Gaussian distribution has fundamental significance in statistics. Many statistical tests such as the t-test require that the variable is question have a Gaussian distribution (Hogg et al., 2005). This distribution is characterized by bell-like shape and an overall symmetry with respect to its peak. The lognormal distribution on the other hand is asymmetric and has much heavier "tail", i.e. decays much slower for large values of the variable than the normal distribution. A surprising number of variables in neuroscience and beyond are described by the lognormal distribution. For example the interspike intervals (Beyer et al., 1975), the psychophysical thresholds for detection of odorants (Devos and Laffort, 1990), the cellular thresholds for detection of visual motion (Britten et al., 1992), the length of words in the English language (Herdan, 1958), and the number of words in a sentence (Williams, 1940) are all united by the fact that their distributions are close to lognormal.

The present results were motivated by the observation that both spontaneous firing rates and synaptic strengths in cortical networks are distributed approximately lognormally. The lognormality of connection strengths was revealed in the course of systematic simultaneous recordings of connected neurons in cortical slices (Song et al., 2005). The lognormality of spontaneous firing rates was observed by monitoring single unit activity in auditory cortex of awake head-fixed rats (Hromadka et al., 2008) using cell attached method. In the traditional extracellular methods cell isolation itself depends upon the spontaneous firing rate: cells with low firing rate are less likely to be detected. During cell attached recordings, cell isolation is independent on the spontaneous or evoked firing rate. Thus cell attached recordings with glass micropipettes permit a relatively unbiased sampling of neurons.

## Novel Hebbian plasticity mechanism

Spontaneous neuronal activity levels and synaptic strengths are related to each other through mechanisms of synaptic plasticity and network dynamics. We therefore asked the question of how could lognormal distributions of these quantities emerge spontaneously in the recurrent network? The mechanism that induces changes in synaptic connectivity is thought to conform to the general idea of Hebbian rule. The specifics of the quantitative implementation of the Hebbian plasticity mechanism are not clear, especially in the cortical networks. Here we propose that a non-linear multiplicative Hebbian mechanism could yield lognormal distribution of connection strengths and spontaneous rates. We propose that the presence of this mechanism can be inferred implicitly from another correlation in the synaptic connectivity matrix. We argued above that the lognormal distribution in spontaneous rates may be produced by correlations between synapses on the same dendrite. By contrast, the signature of the non-linear Hebbian plasticity rule is the presence of similar correlations between synaptic strengths on the same *axon*. Exactly the same test as we proposed to detect dendritic correlations could be applied to axonal data. The presence of both axonal and dendritic correlations leads to the so-called "plaid" connectivity, named so because of both vertical and horizontal correlations present in the synaptic matrix (Figure 5 and 6).

The biological origin of the nonlinear multiplicative plasticity rules is unclear. On one hand, the power-law dependences suggested by our theory [equation (11)] are sublinear in the network parameters, which corresponds to saturation. On the other hand the rate of modification of the synaptic strengths is proportional to the current value of the strength in some power, which is less than one. This result is consistent with the cluster models of synaptic efficacy, in which the uptake of synaptic receptor channels occurs along a perimeter of the cluster of existing receptors (Shouval, 2005). In this case the exponent of synaptic growth is expected to be close to 1/2 [$\beta = 1/2$, see equation (11)].

**Other possibilities**

We have proposed that the lognormal distribution of firing rates emerges from differences in the inputs to neurons. An alternative hypothesis is that the lognormal distribution emerges from differences in the spike generating mechanism that lead to a large variance in neuronal input-output relationship. However, the coefficient of variation of the spontaneous firing rates observed experimentally was almost 120% (Figure 1A). There are no data to suggest that differences in the spike generation mechanism would be of sufficient magnitude to account for such a variance (Higgs et al., 2006).

Another, more intriguing possibility is that the lognormal distribution arises from the modulation of the overall level of synaptic noise (Chance et al., 2002) which can sometimes change neuronal gain by a factor of three or more (Higgs et al., 2006). However, in vivo intracellular recordings reveal that the synaptic input driving spikes in auditory cortex is organized into highly synchronous volleys, or "bumps" (DeWeese and Zador, 2006), so that the neuronal gain in this area is not determined by synaptic noise. Thus modulation of synaptic noise is unlikely to be responsible for the observed lognormal distribution of firing in auditory cortex.

**Conclusions**

The lognormal distribution is widespread in economics, linguistics, and biological systems (Bouchaud and Mezard, 2000; Limpert et al., 2001; Souma, 2002). Many of the lognormal variables are produced by networks of interacting elements. The general principles that lead to the recurrence of lognormal distributions are not clearly understood. Here we suggest that lognormal distributions of both activities and network weights in neocortex could result from specific correlations between connection strengths. We also propose a mechanism based on Hebbian learning rules that can yield these correlations. Finally, we propose a statistical procedure that could reveal both network correlations and Hebb-based mechanisms in experimental data.

# Supplementary material 1

to "Correlated connectivity and the distribution of firing rates in the neocortex"

by Alexei Koulakov, Tomas Hromadka, and Anthony M. Zador

## The emergence of log-normal distribution in neural nets.

### 1. Introduction

The goal of this note is to formulate and address the seeming paradox that emerges in the studies of the distribution of synaptic strengths in the cortex and the distribution of spontaneous rates. The basic findings can be summarized as follows.

(1) The synaptic weights between pairs of cells chosen randomly are described by the log-normal distribution (LND, defined below) (Song et al., 2005).
(2) The spontaneous rates of cells are also distributed log-normally (LN) (Hromadka et al., 2008).

Simplistically, these two facts contradict to each other, because the spontaneous rates in a large network with LN weights distributed randomly and with no correlation are expected to have well-defined values, distributed narrowly, according to the Gaussian distribution. This statement will be addressed below in detail. Thus, if this statement were true, the experimental fact #2 appears to be in conflict with the fact #1. Since the random LN matrix with no correlations between elements appears to contradict these finding, correlations between network weights are expected. We address possible class of correlations that can make these experimental observations consistent with each other. Finally, we propose a non-linear multiplicative learning rule that can yield the proposed correlations.

The note is organized as follows. In Section 2 we describe the properties of the LND that will be useful in the further analysis. In Section 3 we describe the connection between the spontaneous firing rates and the principal eigenvector problem for synaptic weight matrix. In Section 4 we define the random matrices with uncorrelated elements that we call *regular*. In Section 5 we describe the properties of the principal eigenvectors of regular matrices. In this section we formulate the contradiction between two experimental finding listed above. In Section 6 we describe the properties of weight matrices that *do* have correlations between their elements of the type that yields LND for both synaptic weights and spontaneous rates. This section therefore resolves the paradox stated above. In Section 7 we introduce the type of Hebbian learning rules that yield correlations needed to resolve the paradox. Section 8 lists some motivations for the latter learning rule that make it biologically plausible. Finally in Section 9 we solve the equations of the learning rules.

## 2. The log-normal distribution

Consider a variable $x > 0$ whose logarithm $\xi = \ln x$ has a normal distribution, i.e.

$$\rho(\xi) = \frac{1}{\sqrt{2\pi\sigma^2}} e^{-(\xi-\xi_0)^2/2\sigma^2}, \tag{1}$$

where $\sigma$ and $\xi_0$ are the standard deviation and the mean respectively. The distribution function of $x$ is obtained by assuming $\rho(x)dx = \rho(\xi)d\xi$ that leads to

$$\rho(x) = \rho[\xi(x)]\frac{d\xi(x)}{dx} = \frac{1}{x\sqrt{2\pi\sigma^2}} e^{-[\ln(x/x_0)]^2/2\sigma^2}, \tag{2}$$

where $x_0 = e^{\xi_0}$. The probability distribution (2) is called LND. By changing variables to $\xi$ it is easy to calculate various moments of this distribution i.e.

$$\overline{x^n} \equiv \int_0^\infty x^n \rho(x)dx = x_0^n e^{\sigma^2 n^2/2}. \tag{3}$$

Important for us will be the first and the second moments:

$$\overline{x} = x_0 e^{\sigma^2/2} \tag{4}$$

and

$$\overline{x^2} = x_0^2 e^{2\sigma^2}. \tag{5}$$

The variance of the distribution (also called dispersion) is

$$D(x) = \overline{x^2} - \left(\overline{x}\right)^2 = x_0^2 e^{\sigma^2}\left(e^{\sigma^2} - 1\right) \tag{6}$$

It grows exponentially with increasing $\sigma$.

## 3. The spontaneous activity

We adopt here the simplest model for the network dynamics that is described by linear equations

$$f(t + \Delta t) = Wf(t) + i(t) \qquad (7)$$

Here $f(t)$ is the column-vector describing the firing rates of $N$ neurons in the network at time $t$. The input vector $i(t)$ represents the external inputs. The square weight matrix $W$ describes the synaptic weights in the system.

In the absence of synaptic inputs we obtain

$$f(t + \Delta t) = Wf(t). \qquad (8)$$

Spontaneous firing rate is defined here as the average over time firing rate in the absence of external inputs:

$$f \equiv \overline{f(t)} \qquad (9)$$

Spontaneous firing rate is therefore a right eigenvector of the synaptic weight matrix with the eigenvalue equal to one

$$f = Wf \qquad (10)$$

It is therefore the eigenvector that does not decay over time. The other eigenvectors of $W$ are expected to decay as a function of time. They are expected to have the eigenvalues whose absolute values are less that one.

Using another method one can motivate taking the principal eigenvalue of the weight matrix as the representation of spontaneous activity even when the external inputs cannot be neglected. Indeed, let us average equation (7) over time

$$f = Wf + i. \qquad (11)$$

Here $i$ is the averaged input into the network. Consider the set of right eigenvectors of matrix $W$ that we denote $\vec{\xi}_\alpha$:

$$\sum_n W_{kn} \xi_{\alpha n} = \lambda_\alpha \xi_{\alpha k}. \qquad (12)$$

Using this definition one can solve equation (11) for the vector of spontaneous activities $\vec{f}$:

$$f_n = \sum_{\alpha\beta} \frac{\xi_{\alpha n}}{1 - \lambda_\alpha} (G^{-1})_{\alpha\beta} \sum_k \xi^*_{\beta k} i_k. \qquad (13)$$

Here $G_{\alpha\beta} = \sum_n \xi^*_{\alpha n} \xi_{\beta n}$ is the Gram matrix.

Clearly if one of the eigenvalues, say $\lambda_\alpha$, approaches one, the term in the sum (13) corresponding to this eigenvalue will dominate the solution thus yielding

$$f_n \approx C\xi_{\alpha n}, \tag{14}$$

where $C$ is some constant. Thus in the case when recurrent connections have sufficient strength so that one of the eigenvalues of the weight matrix is close to unity, the corresponding eigenvector represents the spontaneous activities in the network.

## 4. Regular matrices

Consider a square $N$ by $N$ matrix $W$. Consider an ensemble of matrices such that all matrix elements are random numbers that are produced from the same distribution. In addition assume that there are no correlations between different elements. This ensemble of matrices belongs to the class of *regular* matrixes. A more accurate definition of this class is given below. Here we will mention that regular matrices have an eigenvalue that in the limit of large $N$ is much larger than other eigenvalues. Also, the eigenvector corresponding to this eigenvalue has elements that are very close to a constant in the limit of large $N$. This statement is true for an arbitrary distribution of the elements of the matrix. Regular matrices represent therefore the simplest class of random matrices with no correlations. They cannot yield a log-normal distribution of the eigenvector elements. Some other form of random matrices is therefore needed to satisfy both of the requirements postulated in the Introduction.

### Definition: *Regular Matrices*

Consider an ensemble of square matrices $W_{ij}$ of different sizes, from one by one to infinity. This ensemble belongs to the class of regular matrices if the following four requirements are met

(i) The distribution of the matrix elements $\rho(W)$ is the same for every position in the matrices of the same size (assumption of uniformity).

(ii) The distribution of matrix elements is the same for matrices of different sizes in the ensemble, up to maybe a scaling factor. More precisely, for every $N_1$ and $N_2$ describing two different sizes of matrices in the ensemble, there exists a positive constant $C$ such that $\rho_{N_1}(W) = C\rho_{N_2}(CW)$, where $\rho_{N_1}$ and $\rho_{N_2}$ are the distributions of elements of matrices of sizes $N_1$ and $N_2$.

(iii) Matrix elements in different columns are statistically independent. This implies that for any $i$ and $k$

$$\rho(W_{ij}, W_{km}) = \rho(W_{ij})\rho(W_{km}) \tag{15}$$

if $j \neq m$, i.e. the matrix elements belong to different columns.

(iv) The matrix elements are positive on average, i.e.

$$\overline{W_{ij}} > 0 \tag{16}$$

We define the in-degree of the matrix as

$$d_i = \sum_j W_{ij}.  \qquad (17)$$

Define $\bar{d}$ and $\sigma(d)$ the average and the standard deviation of the in-degrees for the ensemble. Property (iv) in the definition of regular matrices leads immediately to

$$\bar{d} > 0 \qquad (18)$$

It can be also be shown easily that due to central limit theorem and independence of elements in columns the coefficient of variation of in-degrees becomes infinitely small for an increasing size of the matrix, i.e. when $N \to \infty$

$$\frac{\sigma(d)}{\bar{d}} \equiv \varepsilon \equiv \frac{1}{\bar{d}N} \overline{\sum_i (d_i - \bar{d})^2} \to 0 \qquad (19)$$

Smallness of the coefficient of variation is at the basis of perturbation theory used in this supplement.

**Example 1: Binary Matrices**

$W_{ij} = 0$ or $1$. Assume that $p(W_{ij} = 1) = s$. The number $s \leq 1$ is therefore the sparseness of the matrix. Assume that no correlations are present among matrix elements. For the average in-degree and the standard deviation we obtain after simple calculation

$$\bar{d} = sN \qquad (20)$$

and

$$\sigma^2(d) = Ns(1-s). \qquad (21)$$

Parameter $\varepsilon$ defined in (19) is then

$$\varepsilon = \sqrt{\frac{1-s}{Ns}} \propto \frac{1}{N^{1/2}} \to 0 \qquad (22)$$

when $N \to \infty$. Since the CV of in-degrees vanishes for large $N$, the ensemble of such matrices belongs to the class of regular matrices.

**Example 2: White-Noise Matrices.**

Consider random matrices with uncorrelated matrix elements. We will assume that all elements have the same distribution. We call this type of matrices white-noise. Let us consider sparse matrices for which $\rho(w)$ is the conditional probability distribution for non-zero matrix elements. This distribution can be for example LN. The probability to have a non-zero element (sparseness) is defined by $s$ as in the previous example. The CV of the in-degree for these matrices is

$$\varepsilon = \frac{\sigma(d)}{\bar{d}} = \frac{1}{\sqrt{Ns}} \frac{\sqrt{\overline{w^2} - s\bar{w}^2}}{\bar{w}}, \qquad (23)$$

where $\bar{w}$ and $\overline{w^2}$ are the average and average square of the non-zero matrix elements. Since $\varepsilon$ goes to zero in the limit of increasing matrix size this ensemble of matrices also belongs to the class of regular matrices. Equation (22) is a specific case of a more general expression (23). If for example the distribution of non-zero elements $\rho(w)$ is LN, such as (2), the CV of in-degree is

$$\varepsilon = \sqrt{\frac{e^{\sigma^2} - s}{Ns}}, \qquad (24)$$

as follows from equations (4) and (5).

## 5. Principal eigenvector of the regular matrices

Here we will show that the principal eigenvector of the regular matrices has elements that are normally distributed. The CV of this distribution is equal to the parameter $\varepsilon$ introduced by us in the previous section. Since $\varepsilon \to 0$ for large matrices [equation (19)], the elements of the eigenvector that represent the individual firing rates of neurons have Gaussian distribution with vanishing variance. This claim is valid even if the distribution of the matrix elements is LN, since it is true for any regular matrix (see example 2 above). Thus LN distribution of matrix elements in the absence of correlations yields the eigenvector with small variance in the individual elements (firing rates). Thus, experimental observation (2) (LN spontaneous firing rates) cannot follow from observation (1) (LN weights) in the absence of correlations. In the end of this section we discuss what type of correlations can resolve the paradox.

Consider a square $N$ by $N$ regular matrix $W$. That the matrix is regular, according to (16) requires that the average of the matrix element $\bar{W}$ is positive. Note that here by $\bar{W}$ we mean the average of all matrix elements: positive, negative, and equal to zero; whereas above we used the notation $\bar{w}$ for the average non-zero matrix element of a sparse matrix. It is instructive to first approximate $W$ by the constant matrix, i.e. the one that contains the same value $\bar{W}$ at each position. Let us denote such a matrix by $W^{(0)}$:

$$W_{ij}^{(0)} = \bar{W} \text{ for any } i \text{ and } j. \tag{25}$$

The principal eigenvalue and eigenvector of this matrix are easy to guess. Indeed, if $f_i = 1$ for any $i$, its easy to verify that

$$\sum_j W_{ij}^{(0)} f_j = N\bar{W} f_j. \tag{26}$$

Thus a constant vector is an eigenvector of $W_{ij}^{(0)}$ with the eigenvalue equal to $N\bar{W}$. The other eigenvectors are orthogonal to it because $W_{ij}^{(0)}$ is symmetric. Therefore the sum of the elements of these other eigenvectors is zero. Hence their eigenvalues are also zeros. The constant vector is therefore a principal eigenvector of $W_{ij}^{(0)}$, i.e. its corresponding eigenvalue has a maximum absolute value.

We then calculated the principal eigenvector of $W$ using $W^{(0)}$ as the starting point. We used the perturbation theory that is described in section 10. The result that we got for the eigenvector and the eigenvalue are:

$$f_i = 1 + \frac{d_i - \bar{d}}{\bar{d}} \qquad (27)$$

and

$$\lambda = \bar{d} + \frac{1}{N}\sum_i (d_i - \bar{d}) \qquad (28)$$

Here $d_i$ is the in-degree defined by (17). The correction to the eigenvector in (27) is of the order of $\sigma(d)/\bar{d} = \varepsilon \ll 1$ for large regular matrices. Similar statement can be made about the correction to the principal eigenvalue in (28). CV of the in-degrees serves therefore as the 'smallness' parameter in the perturbation theory.

**The paradox formulated**

Our results show that two experimental constraints listed in the introduction cannot be simultaneously satisfied. The distribution of the elements of the eigenvector (27) is normal due to the central limit theorem, as a distribution of sums of independent random variables. The CV of the distribution is equal to the parameter $\varepsilon \ll 1$ for the regular matrices. This result holds even if distribution of the individual matrix elements is LN, since such matrices are also regular (24). Thus it is impossible for regular matrices to have both their matrix elements and the elements of the principal eigenvector to be LN. The latter will be distributed normally, with a small CV. We arrive at the conclusion that cortical connectivity must contain correlations of the type that makes them not regular.

**Example 3: Suffix (Column) Matrices**

Consider a set of matrices that are formed by products of white-noise matrix $A_{ij}$ and the white-noise random vector $v_j$.

$$B_{ij} = A_{ij} v_j \qquad (29)$$

We will assume that all of the elements of the vector are drawn from the same distribution and that they are not correlated with elements of matrix $A_{ij}$. The ensemble of matrices $B_{ij}$ is regular because the matrix elements located in different columns are not correlated. This is despite the presence of correlations between matrix elements in the same column induced by the common multipliers. We will also show that the distribution of the elements of the principal eigenvector is sharper that that of matrix with no correlations.

Consider the in-degree of matrix $B_{ij}$

$$d_i = \sum_j B_{ij} \tag{30}$$

The white-noise correlations between elements of $v_j$ and $A_{ij}$ can be described as follows:

$$\overline{v_i v_j} = \sigma^2(v)\delta_{ij} + \overline{v}^2 \tag{31}$$

$$\overline{A_{ij} A_{km}} = \sigma^2(A)\delta_{ik}\delta_{jm} + \overline{A}^2 \tag{32}$$

These relationships lead to the expression for the cross-correlations between matrix elements of $B_{ij}$.

$$\overline{B_{ij} B_{km}} = \sigma^2(A)\sigma^2(v)\delta_{ik}\delta_{jm} + \sigma^2(A)\overline{v}^2\delta_{ik}\delta_{jm} + \overline{A}^2\sigma^2(v)\delta_{jm} + \overline{A}^2\overline{v}^2 \tag{33}$$

Our argument will hinge on the following equation describing correlations between in-degrees which is an immediate consequence of Eq. (33).

$$\overline{d_i d_k} = N\sigma^2(A)\sigma^2(v)\delta_{ik} + N\sigma^2(A)\overline{v}^2\delta_{ik} + N\overline{A}^2\sigma^2(v) + N^2\overline{A}^2\overline{v}^2 \tag{34}$$

Because the average in-degree is

$$\overline{d} = N\overline{A}\overline{v} \tag{35}$$

the coefficient of variation (CV) for the in-degrees is

$$\varepsilon^2 = \frac{\overline{d^2} - \overline{d}^2}{\overline{d}^2} = \frac{\sigma^2(A)\sigma^2(v) + \sigma^2(A)\overline{v}^2 + \overline{A}^2\sigma^2(v)}{\overline{A}^2\overline{v}^2} \frac{1}{N} \tag{36}$$

Since $\varepsilon \to 0$ when $N \to \infty$, $B$ is a regular matrix. We also note that

$$\frac{\sigma^2(B)}{\overline{B}^2} = \frac{\sigma^2(A)}{\overline{A}^2} + \frac{\sigma^2(v)}{\overline{v}^2} \tag{37}$$

For this reason the expression for $\varepsilon$ can also be rewritten as follows

$$\varepsilon^2 = \frac{\sigma^2(A)\sigma^2(v) + \sigma^2(B)}{\overline{B}^2} \frac{1}{N}, \tag{38}$$

CV for a white noise matrix can be obtained from (38) by assuming that $\sigma^2(v) = 0$.

$$\varepsilon^2 = \frac{\sigma^2(W)}{\overline{W}^2} \frac{1}{N} \tag{39}$$

Therefore CV for a prefix matrix (38) is larger or equal than that of a white noise matrix with the same distribution of individual elements. Because both types of matrices are regular, their principal eigenvector has the elements given by Eq.

(27). Therefore the CVs of eigenvector elements and in-degrees are the same. Thus expressions (38) and (39) can be understood as the CVs of the eigenvector elements for these two types of matrices. The conclusion about larger CV of the prefix matrix than that of the white-noise matrix is misleading however because in the case of the prefix matrix there is substantial correlation between eigenvector elements. Because (38) describes variability when averaging includes different matrices it does not reflect these correlations. For example, imagine that $A$ is a constant matrix. In this case the in-degrees will still have some variability when considering an ensemble of matrices of the same size. This variability is described accurately by (38). However is this case *all* of the in-degrees are the same for a single matrix which implies, according to (27), that all of the elements of the principal eigenvector are the same. This means that eigenvector elements have no difference for a single matrix.

To describe the distribution of the principal eigenvector elements in individual matrices we introduce the following measure:

$$\Delta^2 \equiv \overline{\left(d_i - \frac{1}{N}\sum_j d_i\right)^2} \qquad (40)$$

that describes the variance of in-degrees with respect to the mean in-degree calculated for the *same* matrix. Opening the brackets and using (34) we obtain

$$\Delta^2 = (N-1)\sigma^2(A)\overline{v^2}. \qquad (41)$$

Therefore this measure of variance goes to zero when elements of matrix $A$ are all the same yielding no difference between elements of the eigenvector as suggested in the end of the last paragraph. The coefficient of variation for $\Delta$ is

$$\frac{\Delta^2}{\overline{d^2}-\overline{d}^2} = \frac{\sigma^2(A)\sigma^2(v)+\sigma^2(A)\overline{v}^2}{\sigma^2(A)\sigma^2(v)+\sigma^2(A)\overline{v}^2+\overline{A}^2\sigma^2(v)}\frac{N-1}{N}<1. \qquad (42)$$

That this ratio is below one explains our observation made in the main paper that the distribution of spontaneous firing rates (elements of the principal eigenvector) for prefix matrix is narrower than for a white-noise matrix with the same distribution of individual matrix elements.

## 6. Irregular matrices

Consider a regular matrix $A_{ij}$. Consider then another matrix $B_{ij}$ that is produced by multiplying all rows of $A_{ij}$ by the elements of random vector $v_i$ whose mean value is larger than zero:

$$B_{ij} = v_i A_{ij} \tag{43}$$

These matrices can also be called *prefix* or *row* matrices. We assume here that $A_{ij}$ and $v_i$ are not correlated. The matrix $B_{ij}$ may or may not belong to the class of regular. Indeed, the in-degrees of $B_{ij}$ are

$$b_i = \sum_j B_{ij} = v_i d_i, \tag{44}$$

where $d_i$ is the in-degree of the regular matrix $A$. Since the latter are distributed with low CV [cf. (19)] the distribution of $b_i$ is dependent upon the distribution of the elements of vector $v_i$ that we denote $\rho(v)$. If the CV of $\rho(v)$ is small, matrix $B$ is regular. If, on the other hand, the CV of $\rho(v)$ does not vanish in the limit of large matrices ($N \to \infty$) matrix $B$ is not regular. It remains to be seen or proven that any irregular matrix can be decomposed into the product of the form (43). We will not prove or disprove this statement in this note.

### Eigenvectors of irregular matrices

Consider now the eigenvector problem for matrix (43). It is formulated as follows

$$v_i \sum_j A_{ij} f_j = \lambda f_i \tag{45}$$

If one introduces the notation

$$y_i = f_i / v_i \tag{46}$$

the eigenvector equation (45) can be rewritten as follows

$$\sum_j A_{ij} v_j y_j = \lambda y_i \tag{47}$$

Thus $y_i$ is the eigenvector of the matrix $C_{ij} = A_{ij} v_j$. Here $A_{ij}$ is random regular while $v_i$ is the random vector. In Example 3 above we showed that matrix $C_{ij}$ is regular. Since $C_{ij}$ is regular its principal eigenvector is approximately constant, as follows from (27)

$$y_i \approx 1 \tag{48}$$

and

$$f_i \approx v_i \tag{49}$$

This approximate equation becomes more and more precise in the limit $N \to \infty$, as follows from (27). We conclude that the distribution of the components of the principal eigenvector matches that of the outer product vector $v_i$.

**Example 4: Lognormal Irregular Matrices**

Consider a matrix whose element have log-normal distribution. For simplicity we will represent the matrix elements in the exponential form

$$A_{ij} = e^{\xi_{ij}} \tag{50}$$

where $\xi_{ij}$ has a normal distribution. If we assume that all $\xi_{ij}$ are taken from the same distribution and are uncorrelated, matrix $A$ is regular. We will now consider a vector $v_i$, whose components are also LN distributed i.e.

$$v_i = e^{\eta_i} \tag{51}$$

where $\eta_i$ are normally distributed and are not correlated with each other. They are also not correlated with the elements of the matrix $\xi_{ij}$. Let us now construct an irregular matrix $B_{ij}$ using the (43) as a prescription:

$$B_{ij} = v_i A_{ij} = e^{\eta_i + \xi_{ij}}. \tag{52}$$

Because each element of $B$ is an exponential of the sum of two normally distributed quantities, it is LN. Also, according to (49) the principal eigenvector of $B$ is LN distributed:

$$f_i \approx v_i = e^{\eta_i} \tag{53}$$

The approximate equality here becomes asymptotically exact in the limit $N \to \infty$ as commented earlier. Thus we arrive at the matrix for which two statements are true, at least, in the limit $N \to \infty$

(1) The elements of matrix $B$ are LN

(2) The components of its principal eigenvector are LN

Since we have suggested a relation between the eigenvector problem and the spontaneous rates in Section 3, these two features may match the corresponding experimental observations listed in the Introduction. Thus, it is possible that cortical networks and the spontaneous activity are produced by irregular matrices, for example, having strong correlations between the outgoing connections, as suggested here.

## 7. The learning rule

Here we propose the learning rule that can yield the irregular matrices in the final stable state. We propose the non-linear multiplicative learning rule for the recurrent synaptic matrix $W_{ij}$

$$W_{ij}(t+\Delta t) = \varepsilon_1 f_i^\alpha W_{ij}^\beta(t) f_j^\gamma + (1-\varepsilon_2)W_{ij}(t). \tag{54}$$

Here we introduced three exponents $\alpha$, $\beta$, and $\gamma$ that describe the non-linearity. It is reasonable to assume that these exponents are positive. The two constants that describe the rates of modification of the components of the weight matrix are $\varepsilon_1$ and $\varepsilon_2$. The former parameter describes the rate of acquiring the new values, while the latter determines the rate of 'forgetting' of the current values of synaptic strengths. The spontaneous rates of the neurons are contained in the components of the vector $f_i$, which is given by the principal eigenvector of the weight matrix

$$\sum_j W_{ij}(t) f_j = \lambda f_j. \tag{55}$$

We assume that the average value for the spontaneous rates is determined by e.g. metabolic constraints

$$\sum_i f_i / N = \overline{f} \tag{56}$$

The average spontaneous rate $\overline{f}$ is assumed to be constant and independent on time.

Before providing biological motivation for the learning rule (54) in the next section we will show that this rule will yield the irregular matrix of the form (52). To this end we consider the final stationary state described by the condition

$$W_{ij}(t+\Delta t) = W_{ij}(t) \equiv W_{ij} \tag{57}$$

Putting this condition into (54) we obtain for the stationary value $W_{ij}$ the following equation

$$\varepsilon_2 W_{ij} = \varepsilon_1 f_i^\alpha W_{ij}^\beta f_i^\gamma \tag{58}$$

This equation has two solutions:

$$W_{ij} = 0 \tag{59}$$

and

$$W_{ij} = (\varepsilon_1/\varepsilon_2)^{1/(1-\beta)} f_i^{\alpha/(1-\beta)} f_j^{\gamma/(1-\beta)}. \tag{60}$$

Which one of the solutions has to be chosen? From the form of the equation (54) it follows that a weight cannot become zero if originally it was above zero. Conversely a synaptic weight that is zero will remain equal to it forever. Thus, the connectivity matrix is preserved during the process described by (54). An element of connectivity matrix $C_{ij}$ is equal to one if there is a synapse from cell $j$ to cell $i$ and zero otherwise. It is thus equal to the transposed adjacency matrix as defined in the graph theory. Two solutions (59) and (60) can be combined into one formula using the connectivity matrix:

$$W_{ij} = (\varepsilon_1 / \varepsilon_2)^{1/(1-\beta)} f_i^{\alpha/(1-\beta)} f_j^{\gamma/(1-\beta)} C_{ij} \tag{61}$$

The synaptic matrix itself depends on the spontaneous rates in the stationary state, which complicates the solution. From the formulation of the eigenvector problem

$$\sum_j W_{ij} f_j = \lambda f_j \tag{62}$$

we obtain the following equation for $f$

$$f_i = \frac{1}{\lambda^M} \left(\frac{\varepsilon_1}{\varepsilon_2}\right)^{\frac{1}{1-\alpha-\beta}} \left(\sum_j C_{ij} f_j^{\frac{\gamma}{1-\beta}}\right)^M \tag{63}$$

Here

$$M = \frac{1-\alpha}{1-\alpha-\beta} \tag{64}$$

Matrix $C_{ij} f_j^{\frac{\gamma}{1-\beta}}$ is regular. Because of this, the sums in (63) are normally distributed with a low CV. If the in-degree of this matrix is $d_i$

$$d_i = \sum_j C_{ij} f_j^{\frac{\gamma}{1-\beta}} = \bar{d} + \Delta_i \tag{65}$$

The random variable $\Delta_i$ is normally distributed and $\sigma(\Delta_i)/\bar{d} = \varepsilon \ll 1$. For the logarithm of the spontaneous rates we can write

$$\ln f_i = C + M \ln\left(1 + \frac{\Delta_i}{\bar{d}}\right) \tag{66}$$

Here $C$ is some constants. Due to the smallness of parameter $\varepsilon = \sigma(\Delta_i)/\bar{d}$ we can expand the logarithm and write

$$\ln f_i = \tilde{C} + M \frac{\Delta_i}{\bar{d}} \tag{67}$$

The logarithm of the spontaneous rate is therefore normally distributed. Of course the standard deviation of this distribution may be small:

$$\sigma(\ln f) = M\varepsilon \qquad (68)$$

because $\varepsilon \ll 1$. However with $\alpha + \beta \to 1$ the exponent $M$ may become large so that the distribution of $f$ becomes LN. Note that the distribution of the non-zero synaptic weights is also LN as the distribution of the product of LN variables (61) .

## 8. Motivation for the learning rule.

The standard Hebbian learning rule would look like this:

$$W_{ij}(t + \Delta t) = \varepsilon_1 f_i f_j + (1 - \varepsilon_2) W_{ij}(t) \qquad (69)$$

There are several ways in which our rule (54) is more biologically plausible than (69).

(1) The learning rule that we postulated (54) preserves connectivity matrix. This means that the sparse matrix of synaptic weight will remain sparse, with the same connectivity. The learning rule (69) produces a full matrix. Since cortical connectivity is sparse (Song et al., 2005), our rule is more biologically plausible.

(2) Our learning rule suggests that the uptake of proteins controlling the synaptic strength occurs at the rate dependent of the number of existing proteins. This is consistent with the models in which the uptake occurs into spatially localized clusters in PSD, which would make the rate of synapse growth larger for a larger synapse (Shouval, 2005).

## 9. Variance of the log-normal distribution.

Eq. (68) for the variance of the logarithm of the firing rates $\sigma^2(\ln f)$ can be rewritten as follows

$$\sigma^2(\ln f) = M^2 \varepsilon^2 \tag{70}$$

where

$$\varepsilon = \varepsilon\left[\sigma^2(\ln f)\right] \tag{71}$$

is the CV of the in-degrees of matrix $C_{ij} f_j^{\frac{\gamma}{1-\beta}}$. Because the matrix itself depends on $f$ the CV of in-degrees is determined by the distribution of the components $f_i$, which is emphasized by the last equation. The exact form of dependence in (71) is easy to derive using (24)

$$\varepsilon^2 = \frac{e^{\sigma^2(\ln f)\left(\frac{1-\beta}{\gamma}\right)^2} - s}{Ns} \tag{72}$$

where $s$ is the sparseness of the connectivity matrix $C_{ij}$, which is by definition

$$s \equiv \frac{1}{N^2} \sum_{ij} C_{ij} \tag{73}$$

i.e. the fraction of its non-zero elements. The full form of the equation which determines $\sigma^2(\ln f)$ is

$$\sigma^2(\ln f) = \frac{M^2}{Ns}\left[e^{\sigma^2(\ln f)\left(\frac{1-\beta}{\gamma}\right)^2} - s\right] \tag{74}$$

This equation should be solved iteratively to find the variance of the logarithm of the firing rates. The solution becomes large when $M = (1-\beta)/(1-\alpha-\beta) \to \infty$, i.e. when $\alpha + \beta \to 1$.

The variance of the logarithm of non-zero weight matrix elements is then given by

$$\sigma^2(\ln W) = \sigma^2(\ln f)\frac{\alpha^2 + \gamma^2}{(1-\beta)^2} \tag{75}$$

The latter relationship is found from (61).

# 10. Perturbation theory solution for the principal eigenvector of the regular matrix.

Here we will prove equations (27) and (28) that are used to demonstrate the smallness in the variation of the components of the principal eigenvector of the regular matrices. To this end we represent a regular matrix $W_{ij}$ as a sum of a constant matrix $W_{ij}^{(0)} = \overline{W} > 0$, whose elements are all the same, and the correction $\Delta W_{ij}$

$$W_{ij} = W_{ij}^{(0)} + \Delta W_{ij} \tag{76}$$

This equation may be understood as the definition of the correction matrix $\Delta W_{ij}$. Despite the fact that the individual elements of $\Delta W_{ij}$ are large, we will assume that the effects of adding this correction on the eigenvector and eigenvalue are small. We will show that this actually is true in the end of calculation. This may be viewed as a circular argument. Indeed, to obtain smallness of the correction we assume that the correction is small. However, we know that there is only one solution. The uniqueness of the solution is provided by the Perron-Frobenius theorem for non-negative matrices. Therefore, obtaining solution that is self-consistent, i.e. does not contradict to itself, is sufficient.

To perform the perturbation theory analysis we will represent the principal eigenvector of the matrix $W_{ij}$ as a sum of the solution of the 'unperturbed' problem $f_i^{(0)}$ and the correction

$$f_i = f_i^{(0)} + \delta_i \tag{77}$$

where $f_i^{(0)} = const$ as we argued before, and the small correction $\delta_i \ll f_i^{(0)}$. The correction can always be made perpendicular to $f_i^{(0)}$ by including the non-perpendicular component of $\delta_i$ into $f_i^{(0)}$. Since $f_i^{(0)} = const$ we conclude that

$$\sum_j \delta_j = 0 \tag{78}$$

The vector $f_i^{(0)}$ is the solution to the 'unperturbed' eigenvector problem

$$\sum_j W_{ij}^{(0)} f_j^{(0)} = \lambda^{(0)} f_i^{(0)}, \tag{79}$$

where, according to (26)

$$\lambda^{(0)} = N\overline{W}. \tag{80}$$

The vector $f_i$ is the solution to the full problem

$$\sum_j W_{ij} f_j = \lambda\, f_i \qquad (81)$$

where

$$\lambda = \lambda^{(0)} + \Delta\lambda \qquad (82)$$

Equations (77), (81), and (82) can be combined as follows

$$\sum_j (W_{ij}^{(0)} + \Delta W_{ij})(f_j^{(0)} + \delta_j) = (\lambda^{(0)} + \Delta\lambda)(f_i^{(0)} + \delta_i) \qquad (83)$$

In the expanded form this reads:

$$\sum_j W_{ij}^{(0)} f_j^{(0)} + \sum_j W_{ij}^{(0)} \delta_j + \sum_j \Delta W_{ij} f_j^{(0)} + \sum_j \Delta W_{ij} \delta_j =$$
$$= \lambda^{(0)} f_i^{(0)} + \lambda^{(0)} \delta_i + \Delta\lambda f_i^{(0)} + \Delta\lambda \delta_i \qquad (84)$$

The first term in the l.h.s. cancels with the first term in the r.h.s. because of (79). The second term in the l.h.s. is zero, because $W_{ij}^{(0)}$ is a constant matrix and $\delta_j$ satisfies (78). The forth term in l.h.s. is much smaller than the third, and, therefore can be neglected. The same is true about the fourth term in the r.h.s. in comparison with the third term there. We therefore arrive at a much shorter equation:

$$\sum_j \Delta W_{ij} f_j^{(0)} = \lambda^{(0)} \delta_i + \Delta\lambda f_i^{(0)} \qquad (85)$$

This equation is approximate. However, it is asymptotically correct, when $\delta_i \ll f_i^{(0)}$.

We will now multiply both sides of the equation by $f_i^{(0)}$ and sum over $i$. Because $\delta_i \perp f_i^{(0)}$ the first term in the r.h.s. gives no contribution. We obtain for the correction to the eigenvalue

$$\Delta\lambda = \sum_{ij} f_i^{(0)} \Delta W_{ij} f_j^{(0)} / \sum_i |f_i^{(0)}|^2 = \frac{\vec{f}^{(0)T} \Delta\hat{W} \vec{f}^{(0)}}{|\vec{f}^{(0)}|^2} \qquad (86)$$

Let us estimate this correction. Because the elements of vector $\vec{f}^{(0)}$ are all the same, we can write

$$\Delta\lambda = \frac{1}{N} \sum_i (d_i - \bar{d}) \qquad (87)$$

where

$$d_i = \sum_j W_{ij} \qquad (88)$$

are the in-degrees of matrix $\hat{W}$. The average over the ensemble value of the correction is zero. The variance of the correction is

$$\overline{\Delta\lambda^2} = \frac{1}{N} \overline{(d_i - \bar{d})^2} = \frac{\sigma^2(d)}{N} \qquad (89)$$

The relative correction to the eigenvalue can be estimated to be

$$\frac{\Delta\lambda}{\lambda^{(0)}} = \frac{\sigma(d)}{\bar{d}}\frac{1}{\sqrt{N}} = \frac{\varepsilon}{\sqrt{N}} \ll 1 \tag{90}$$

i.e. is small because both the CV of the in-degrees of the regular matrix $\varepsilon$ is small and the matrix is large.

We will now use the result (90) to find the correction to the eigenvector. Let us estimate various terms in equation (85) that we will recite here for convenience

$$\sum_j \Delta W_{ij} f_j^{(0)} = \lambda^{(0)} \delta_i + \Delta\lambda f_i^{(0)} \tag{91}$$

The first term is of the order of $\sigma(d) f_j^{(0)}$ while the last term in the r.h.s. is equal to

$$\lambda^{(0)} \frac{\varepsilon}{\sqrt{N}} f_i^{(0)} = \bar{d}\frac{\sigma(d)}{\bar{d}}\frac{1}{\sqrt{N}} f_i^{(0)} = \frac{\sigma(d)}{\sqrt{N}} f_i^{(0)} \tag{92}$$

The last term in (91) therefore can be neglected. For the correction to the eigenvector we obtain

$$\delta_i = \frac{1}{\lambda^{(0)}} f_i^{(0)} \sum_j \Delta W_{ij} = f_i^{(0)} \frac{1}{\bar{d}}(d_i - \bar{d}), \tag{93}$$

which is the same as equation (27). Equation (93) also implies that

$$\frac{\delta_i}{f_i^{(0)}} \sim \frac{\sigma(d)}{\bar{d}} \sim \varepsilon \ll 1 \tag{94}$$

i.e. correction to the eigenvector is small.

We now have to show that the neglected term in equation (84), i.e. the fourth term in the l.h.s. is much smaller than the third term in the limit of large matrices

$$s_3 \equiv \sum_j \Delta W_{ij} f_j^{(0)} \gg s_4 \equiv \sum_j \Delta W_{ij} \delta_j \tag{95}$$

Because the expectation value for $s_3$ is zero while $s_4$ may be positive on average we will compare their squares. We obtain

$$\overline{s_3^2} = \sum_{jk} \overline{\Delta W_{ij} \Delta W_{ik}} = N\sigma^2(W) \sim N \tag{96}$$

In deriving this we used property (iii) in the definition of regular matrices (statistical independence of elements in different columns) which leads to

$$\overline{\Delta W_{ij} \Delta W_{ik}} = \sigma^2(W)\delta_{jk} \tag{97}$$

We also assumed that $f_j^{(0)} = 1 \sim \sigma(W)$ for simplicity.

Estimation of $s_4$ requires somewhat larger effort. Using (93) we can write

$$\overline{s_4^2} \equiv \overline{\left(\sum_j \Delta W_{ij}\delta_j\right)^2} \approx \frac{1}{\overline{d}^2}\sum_{jklm}\overline{\Delta W_{ij}\Delta W_{jk}\Delta W_{il}\Delta W_{lm}} \,. \tag{98}$$

Because $\overline{\Delta W_{ij}} = 0$ and elements in different columns (with different second indices) are independent, the sum in this equation breaks into the sum of products of pairs:

$$\overline{s_4^2} \approx \frac{1}{\overline{d}^2}\left[\sum_{\substack{jl \\ j\neq l}}\overline{\Delta W_{ij}\Delta W_{jj}}\cdot\overline{\Delta W_{il}\Delta W_{ll}} + \sum_{\substack{jk \\ j\neq k}}\overline{\Delta W_{ij}\Delta W_{jj}}\cdot\overline{\Delta W_{jk}\Delta W_{jk}} + \right. $$
$$\left. +\sum_{\substack{kj \\ j\neq k}}\overline{\Delta W_{ij}\Delta W_{kj}}\cdot\overline{\Delta W_{jk}\Delta W_{ik}} + \sum_j \overline{\Delta W_{ij}\Delta W_{jj}\Delta W_{ij}\Delta W_{jj}}\right] \tag{99}$$

Because the largest sums in this equation include $N^2$ terms we can estimate $\overline{s_4^2}$ as follows

$$\overline{s_4^2} \sim \frac{N^2}{\overline{d}^2} \sim 1 \ll \overline{s_3^2} \sim N \,. \tag{100}$$

Thus the forth term in equation (84) is much smaller than the third term on average for very large matrices.

# Supplementary material 2

**to "Correlated connectivity and the distribution of firing rates in the neocortex"**

by Alexei Koulakov, Tomas Hromadka, and Anthony M. Zador

## The effects of exponential input-output relationship in the firing of neurons.

In this supplement we will consider a recurrent network of neurons for which the firing rate is an exponential function of the input current, i.e.

$$f = f_0 e^{I/\lambda} \qquad (101)$$

Here $f_0$ and $\lambda$ are constants. A simple explanation of lognormal spontaneous firing rates would be that the input current $I$ for these neurons has a normal distribution as a result of uncorrelated synaptic strengths of many input synapses. As a result the firing rates, as exponentials of the input current, have lognormal distribution. Here we will show that the hypothesis of exponential input-output relationship cannot yield large variance in the logarithm of firing rates for the recurrent network of neurons. We will show that large variance in the logarithm will have to lead to instability in the recurrent network of such neurons. This is based on the extremely strong positive gain in the recurrent network provided by the exponential input-output relationship (101).

We will start by deriving the stability condition for the recurrent network. We will see below that the stability condition cannot be satisfied when the standard deviation of the logarithm of the spontaneous firing rates is substantial, i.e. is close to 1 as required by experimental observations. To proceed with the analysis of stability we introduce the variables for firing rates and weights

$$f_i = e^{\xi_i} \qquad (102)$$

$$w_{ij} = e^{\eta_{ij}} \qquad (103)$$

Here indexes $i$ and $j$ label neurons in the networks. The stability condition can especially easily be derived in the case when all neurons have essentially the same firing rates, i.e. $\overline{\delta\xi^2} \ll 1$. Here $\overline{\delta\xi^2}$ is the standard deviation of the logarithm of the firing rates. We will see from this stability condition that it is violated when $\overline{\delta\xi^2} \sim 1$. The latter case is therefore not essential for the stability analysis. Later we will however derive the stability condition for $\overline{\delta\xi^2} \sim 1$ case for the sake of completeness.

**1) The case of small deviations $\overline{\delta\xi^2} \ll 1$.**

For the recurrent current and the variance of the recurrent current we obtain

$$\overline{I} = wsNf + I_0 \tag{104}$$

$$\overline{\delta I^2} = w^2 sNf^2 A \tag{105}$$

Here $w$, $s$, $N$, $I_0$, and $A$ are the average synaptic strength, sparseness of the network, number of neurons, external offset current, and a numerical coefficient of the order of one. For the lognormal distribution of synaptic weights it can be derived that

$$A = e^{\overline{\delta\eta^2}} - s \tag{106}$$

Experimental evidence suggests that $\overline{\delta\eta^2} \approx 1$ for cortical networks. Equations (104) and (105) are typical for the sum of independent random variables in which case both the average and the variance are proportional to the number of terms in the sum, i.e. $N$. The variance in the logarithm of the firing rates can be related to the variance of recurrent current through the input-output relationship (101)

$$\overline{\delta\xi^2} = \frac{\overline{\delta I^2}}{\lambda^2} = \frac{w^2}{\lambda^2} Nsf^2 A \tag{107}$$

Stability condition for the recurrent network reads

$$\frac{df}{dI} = \frac{f}{\lambda} < \frac{1}{d\overline{I}_{rec}/df} = \frac{1}{wsN} \tag{108}$$

Note that here one can disregard the difference between the current on the input of each neuron and the average current because of the condition $\overline{\delta\xi^2} \ll 1$. The gain in the input-output relationship $\lambda$ can be excluded from the last equation using equation (107). After this substitution we arrive at the final result of this subsection, which expressed by the stability condition of the recurrent network in terms of the parameters of the lognormal distribution

$$\overline{\delta\xi^2} < \frac{A}{Ns}. \tag{109}$$

Therefore, for large networks ($N \gg 1$) stability condition is impossible to satisfy if the logarithm of the firing rates has substantial variance i.e. $\overline{\delta\xi^2} \sim 1$. Experimental observations of large variance of the logarithm are therefore hard to reconcile with the exponential input-output relationship (101).

## 2) The case of substantial variance $\overline{\delta\xi^2} \sim 1$.

We will argue here that condition similar to (109) has to be satisfied in this case as well. The exact form of the condition is

$$\overline{\delta\xi^2} < \frac{e^{\overline{\delta\eta^2}+\overline{\delta\xi^2}} - s}{Ns} \tag{110}$$

It is possible to satisfy this inequality if $\overline{\delta\xi^2} \ll 1$ and if $\overline{\delta\xi^2} \approx \ln Ns \gg 1$. We note however that the latter case is not consistent with experiments in which $\overline{\delta\xi^2} \approx 1$ is observed.

Our analysis is essentially based on the following equations that can be easily confirmed for lognormal variables

$$\overline{f} = \overline{e^\xi} = e^{\overline{\xi}+\overline{\delta\xi^2}/2}. \tag{111}$$

$$\overline{\delta f^2} = e^{2\overline{\xi}+\overline{\delta\xi^2}}(e^{\overline{\delta\xi^2}} - 1) \tag{112}$$

The coupled dynamics of the network current and the variance of the firing rates can be described by the following equations

$$\overline{\delta\xi^2}(t+1) = \frac{Ns}{\lambda^2} e^{2\overline{\eta}+2\overline{\xi(t)}} e^{\overline{\delta\eta^2}+\overline{\delta\xi^2}(t)} (e^{\overline{\delta\eta^2}+\overline{\delta\xi^2}(t)} - s) \tag{113}$$

$$\overline{I(t+1)} = Ns e^{\overline{\eta}+\overline{\xi(t)}} e^{\overline{\delta\eta^2}/2+\overline{\delta\xi^2}(t)/2} + I_0 \tag{114}$$

Here we assumed that the network weights described by the variables $\eta$ do not change with time and the weight matrix is uncorrelated. In the equilibrium we have

$$\overline{\delta\xi_0^2} = \frac{Ns}{\lambda^2} e^{2\overline{\eta}+2\overline{\xi}} e^{\overline{\delta\eta^2}+\overline{\delta\xi_0^2}} (e^{\overline{\delta\eta^2}+\overline{\delta\xi_0^2}} - s) \tag{115}$$

Dividing (113) by (115) we obtain

$$\frac{\overline{\delta\xi^2}(t+1)}{\overline{\delta\xi_0^2}} = e^{2\overline{\xi(t)}-2\overline{\xi}} e^{\overline{\delta\xi^2}(t)-\overline{\delta\xi_0^2}} \frac{e^{\overline{\delta\eta^2}+\overline{\delta\xi^2}(t)} - s}{e^{\overline{\delta\eta^2}+\overline{\delta\xi_0^2}} - s} \tag{116}$$

Using the relationship

$$\overline{\xi(t)} = \overline{I(t)}/\lambda + \ln f_0 \tag{117}$$

and introducing small deviations from the equilibrium

$$\overline{\delta\xi^2}(t) = \overline{\delta\xi_0^2} + \Delta(t) \tag{118}$$

$$\overline{I(t)} = \overline{I} + \Delta I(t) \tag{119}$$

we obtain the following linear system of equations for the small deviations

$$\Delta I(t+1) = \kappa \Delta I(t) + \frac{\kappa\lambda}{2}\Delta(t) \qquad (120)$$

and

$$\Delta(t+1) = \overline{\delta\xi_0^2}\frac{2}{\lambda}\Delta I(t) + \overline{\delta\xi_0^2}\frac{2e^{\overline{\delta\eta^2}+\overline{\delta\xi_0^2}}-s}{e^{\overline{\delta\eta^2}+\overline{\delta\xi_0^2}}-s}\Delta(t). \qquad (121)$$

Here the coefficient

$$\kappa = \sqrt{\frac{\overline{\delta\xi_0^2}Ns}{e^{\overline{\delta\eta^2}+\overline{\delta\xi_0^2}}-s}}. \qquad (122)$$

It can be shown that in the large-$N$ limit the eigenvalues of the system (120) and (121) are below 1 in absolute value if $\kappa<1$, i.e. when condition (110) is satisfied. Because $\overline{\delta\xi_0^2} \equiv \overline{\delta\xi^2} \approx 1$ experimentally and $Ns \gg 1$ the condition $\kappa<1$ is difficult to satisfy. The hypothesis of exponential firing rates (101) is therefore not compatible with the lognormal distribution produced by the recurrent networks because of the lack of stability.